\documentclass[draftcls,onecolumn, 11 pt]{IEEEtranA}
\usepackage{graphicx}
\usepackage{multirow}
\usepackage{amssymb}
\usepackage[cmex10]{amsmath}
\usepackage{cite}
\usepackage{subfig}
\usepackage{gensymb}
\usepackage[T1]{fontenc}
\usepackage{mathptmx}
\usepackage[scaled]{helvet}    
\usepackage{luximono}          
\usepackage{txfonts}

\def\cB{{\cal B}}

\def\tt{\tilde t}

\newcount\exnum

\long\def\bexer{%
\global\advance\exnum by1
\global\insertnum=\exnum
\binsert{Exercise}}

\begin{document}

\title{TCP-like molecular communications}

\author{Luca~Felicetti, Mauro~Femminella$^*$,~\IEEEmembership{Member,~IEEE}, Gianluca~Reali,~\IEEEmembership{Member,~IEEE}, Tadashi~Nakano,~\IEEEmembership{Member,~IEEE}, Athanasios V. Vasilakos, \IEEEmembership{Senior Member,~IEEE}%
\thanks{L. Felicetti, M. Femminella, and G. Reali are with the Department of Engineering, University of Perugia, via G. Duranti 93, 06125 Perugia, Italy, (email: \{mauro.femminella,gianluca.reali\}@unipg.it, ing.luca.felicetti@gmail.com). T. Nakano is with the Graduate School of Frontier Biosciences, Osaka University, 1-3 Yamadaoka, Suita, Osaka 565-0871, Japan (email: tadasi.nakano@fbs.osaka-u.ac.jp). A. V. Vasilakos is with the Department of Computer and Telecommunications, Engineering, University of Western Macedonia, Macedonia, Greece (email: vasilako@ath.forthnet.gr). Asterisk indicates corresponding author.}}    
    
\thispagestyle{empty}

\IEEEaftertitletext{\vspace{-2.4\baselineskip}\noindent\begin{abstract} 
In this paper, we present a communication protocol between a pair of biological nanomachines, transmitter and receiver, built upon molecular communications in an aqueous environment. In our proposal, the receiver, acting as a control node, sends a connection setup signal to the transmitter, which stokes molecules, to start molecule transmission. The molecules transmitted by the transmitter propagate in the environment and are absorbed by the receiver through its receptors. When the receiver absorbs the desired quantity of molecules, it releases a tear-down signal to notify the transmitter to stop the transmission. The proposed protocol implements a bidirectional communication by using a number of techniques originally designed for the TCP. In fact, the proposed protocol is connection-oriented, and uses the TCP-like probing to find a suitable transmission rate between transmitter and receiver so as to avoid receiver congestion. Unlike the TCP, however, explicit acknowledgments are not used, since they would degrade the communication throughput due to the large delay, a characteristic feature of molecular communications. Thus, the proposed protocol uses implicit acknowledgments, and feedback signals are sent by the receiver to throttle the transmission rate at the transmitter, i.e., explicit negative feedbacks. We also present the results of an extensive simulation campaign, used to validate the proposed protocol and to properly dimension the main protocol parameters.
\end{abstract}
\noindent
\vspace{1\baselineskip}}

\maketitle

\section{Introduction}\label{intro}

Molecular communication is a novel paradigm for communication between biological nanomachines (or bio-nanomachines) over a short range \cite{Akyildiz08,Nakano12c,Nakanobook2013} in an aqueous environment. It consists of the emission and reception of molecules that act as communication signals. Bio-nanomachines are made of biological materials and perform communication and other tasks, such as moving in the environment, sensing a specific type of molecules in the environment, or catalyzing specific biochemical reactions. The size of individual bio-nanomachines can be up to tens of $\mathrm{\mu m}$, and their capabilities are strictly limited by their size. Thus, interaction of bio-nanomachines through molecular communication is necessary to accomplish complex tasks. The simplest form of molecular communication consists of a group of sender bio-nanomachines which transmit a burst of molecules, the molecules propagate in the environment by diffusion, and a group of receiver bio-nanomachines chemically react with the molecules, so receiving a chemical signal. Since bio-nanomachines are made of biological materials, molecular communication provides a simple yet effective mechanism for bio-nanomachines to communicate, without the need of integrating components for electromagnetic communications into them.

This paper proposes a complete molecular communication protocol for bio-nanomachines at the nanoscale. We consider a simple yet meaningful scenario consisting of two fixed bio-nanomachines, one acting as receiver and control node (RX), the other as transmitting node (TX). The communication happens using two different types of molecules, which propagate by diffusion, modeled as Brownian motion \cite{Philibert2006}. Those transmitted by the TX node and representing the signal to be delivered are labeled as \textit{S}, those transmitted by the control node RX to encode control messages are labeled \textit{R}. The control signals are encoded in different patterns of bursts of the \textit{R} molecules by using the on-off keying modulation. Timing and reception threshold have been defined in order to ensure applicability of \textit{the same values} on a range of communication distances. Our aim is to identify suitable system parameters so as to allow proper system operation for communication ranges between 20 and 70 $\mu m$. This aspect is crucial for the deployment of a molecular communication system, where the communication range may be estimated but not be exactly determined in advance. Consequently, the size of control bursts is not fixed, but depends on the distance between TX and RX node, and the burst size is autonomously determined by the RX through an adaptive ranging algorithm, borrowed by the WiMAX protocol \cite{WiMAX}.

Differently from most of the works in this field (see Section \ref{back}), which deals almost exclusively with physical layer issues, motivated by the fact that this research field is extremely new and physical layer communications are still not consolidated, this work proposes a connection oriented mechanism. This protocol builds upon a substantial body of research performed by other works (\cite{Pierobon10, Kuran201086,6612809,Moore12a}), and extends the pioneering work carried out in \cite{nakano13}, which explored rate control mechanisms in molecular communication. In computer networks, most of application protocols make use of the transmission control protocol (TCP) \cite{Tanenbaum02}, a bidirectional network protocol operating at the transport layer which provides reliability, in-sequence delivery, congestion control, and flow control to applications running in network endpoints. While complete implementation of the TCP in molecular communication is difficult or useless, some of its functions, namely connection oriented capability, reliability, and congestion and flow control, could be translated in such an environment. Connection oriented capabilities allow a control channel to be set up between the sender and the receiver, which will be able to control the molecule release rate at the sender on the basis of the number of molecules absorbed or detected at target nodes in the surrounding environment. In fact in molecular communication, if the sender bio-nanomachine keeps transmitting molecules, the number of molecules in the environment increases. Since the receiver bio-nanomachines are able to react at a limited rate, the molecules remaining in the environment eventually degrade and result in loss of molecules. Transmission rate needs to be adjusted to reduce the loss rate in specific applications where the molecules are expensive, limited in number \cite{Freitas1988} or where lost molecules may cause undesired side effects. Since typical delays in molecular communications are much larger than those in computer networks, a solution based on explicit feedback is not suitable, since it would increase communication delay. Further, the analysis carried out in \cite{nakano13} suggests that negative feedbacks, when interpreted by the recipient as the indication to reduce the transmission rate, can increase the efficiency (or decrease the loss rate) in molecular communication.

In summary, the commonalities between our proposal and TCP consist of:
\begin{itemize}
\item Connection oriented protocol operation: as in TCP, our protocol establishes and tears down a communication session.
\item ``Reliable'' data transfer: as in the TCP, where the target is to transfer a given amount of data reliably, in our case the protocol allows delivering a given amount of molecules in a reliable way, ensuring that the target number of molecules is reached. In other words, reliable data is mapped into reliable transfer of a given number of molecules. Once the delivery of molecules is completed, a stop signal is sent from the receiver to tear down the connection.
\item Receiver driven flow control: as in TCP, the receiver, when congested, adopts strategy to avoid further transmission of molecules.
\item Adaptive transmission bandwidth: we adopted the network probing feature designed for TCP to calculate a suitable transmission rate of molecules, which clearly depends on the distance between the sender and receiver. In more detail, once that the sender has received the start signal from the receiver (connection set up), the sender starts releasing molecules in bursts, and it \textit{linearly} increases the number of molecules transmitted during each burst, in the same manner that TCP does with the number of bytes in the congestion avoidance phase. When a feedback arrives, since it is a negative one, it is interpreted as the indication to reduce the transmission rate. In our protocol, this is implemented by halving the size of the current burst. Then, the transmitter restarts linearly increasing the burst size at each transmission time, exactly as the congestion control mechanism implemented in \textit{TCP Reno} immediately after three duplicated acknowledgments are received \cite{Tanenbaum02}. 
\end{itemize}

Clearly, some TCP features cannot be implemented, such as byte oriented transmissions, encapsulation rules and segment organization. In fact, our protocol does not deliver bit-encoded information from TX to RX, but a given number of molecules. In addition, the TX does not perform any estimation of the RTT, and adopts a fixed transmission window for releasing bursts of molecules.

One possible application of molecular communication, which could highly benefit of our proposal, is the drug delivery, which is of great interest for the medical area. Drug molecules can be carried by bio-nanomachines such as natural cells (e.g., blood cells) or synthetic counterparts \cite{Allen04,Yoo11}. Due to their size, bio-nanomachines can be directly injected close to the target site of drug delivery in the body of a patient. The advantages of drug delivery over conventional drug administration include the potential reduction of side effects by releasing drug molecules only very close to the target site, so prolonging the efficacy of drug molecules through a sustained drug administration in a patient body, while limiting the side effects on healthy cells which are not strictly close to the target. Molecular communication provides novel methodologies for drug delivery by allowing groups of bio-nanomachines to cooperate in order to maximize the therapeutic effect of drug molecules \cite{Nakano12e,Atakan12,Bogdan12}. For instance, bio-nanomachines capable of detecting a target site (e.g., tumor cells) may transmit bursts of molecules to indicate the location of the target site. Other bio-nanomachines with actuation functions, upon detecting these signals, move toward the target site and deliver the specific drug molecules requested, thus improving the targeting accuracy \cite{Maltzahn11}. Beyond the recruitment signal, different bio-nanomachines at the target site may also communicate to perform more complex tasks. One of such tasks is to adjust the rate of drug delivery depending on environmental conditions, such as the spatial distribution of target cells, the distance to actuator bio-nanomachines, the rate of drug uptake by target cells, the rate of drug release by bio-nanomachines, and the amount of drug which has to be delivered to the target.

The paper is organized as follows. In Section \ref{back} we illustrate the related work in the field. The complete protocol design, including physical and higher layer issues, such as control algorithms and protocol state machines, which is the main body of this paper, is presented in Section \ref{design}. The result of the simulation campaign, used to validate the proposed protocol and to analyze the trade-off between throughput and communication efficiency, are presented in Section \ref{perf}. Finally, we draw our conclusions in Section \ref{conc}.

\section{Background}\label{back}

Major efforts in the  area of molecular communication are focused on physical layer issues of various types of molecular communication media. In these efforts, information capacity and physical characteristics (e.g., delay, signal attenuation, amplification, and energy requirements) of molecular communication are studied using random walk models \cite{Moore09,Eckford07a,Nakano12d,Kuran201086}, random walk models with drift \cite{Kadloor2012,Srinivas12}, diffusion-based models \cite{Akan08,Atakan10,Mahfuz10,Pierobon10,Pierobon11c}, diffusion-reaction-based models \cite{Nakano10a,Nakano11b}, active transport models \cite{Moore09,Eckford10}, and a collision-based model \cite{Guney2012}. 

As for diffusion-based models, a review of different transmission schemes is provided in \cite{ShahMohammadian2012}, where the authors classify existing schemes into pulse position modulation (PPM, \cite{Kadloor2012}) or concentration shift keying (CSK, \cite{Atakan2010}), and propose molecule shift key (MoSK), which uses a combination of bursts of different molecules to encode signals. The same authors also present a study about synchronization for MoSK \cite{shahmo}, identifying the issues derived by the use of different molecules and a possible solution. The possible trade-off between symbol duration and communication distance is analyzed in \cite{6612809}. 

In addition, recent efforts address higher layer and other important issues in molecular communication. For instance, in \cite{Moore11d}, the beacon coordinate system implements an addressing mechanisms in molecular communication. Considerations about complexity and scope of nanonetworks have revealed the need for synchronizing several nanomachines. 
In \cite{Moore11c}, diffusion-based mechanisms of synchronization are designed to allow bio-nanomachines to coordinate the timing of their actions through the use of inhibitory molecules. In \cite{Moore12a}, distance measurement protocols are developed for a bio-nanomachine to measure distance to another bio-nanomachine by monitoring the patterns of propagating molecules (e.g., round-trip-time, amplitude fading). In \cite{Cobo10}, a routing system is presented in which a sender bio-nanomachine transmits information molecules using a mobile carrier with addressing molecules to indicate the desired receiver bio-nanomachines, and router bio-nanomachines, upon receiving the mobile carrier, apply specific chemical processes to retransmit the mobile carrier to the next hop router bio-nanomachine on the path to the receiver bio-nanomachine.

Further, a recent effort describes a complete view of a layered network architecture of molecular communication \cite{Vasilakos14}. Following the layered architecture of traditional communication networks such as the Open Systems Interconnection model (OSI) and TCP/IP reference model, it develops a formal model for each layer, explains how each layer behaves, and identifies potential research directions for each layer. One of the key research issues discussed in \cite{Vasilakos14} is to examine the role and methods of feedback in molecular communication, which is to be investigated in the present paper.

Other works deal with simulation software for nano-scale communications. The simulator illustrated in \cite{Gul2010138}, which is based on NS-2, is written in C++ and Tcl. It implements the laws governing a tridimensional (3D) Brownian motion and the multiparticle lattice gas automata algorithm, which consists of partitioning the propagation medium into lattice sides. Particle positions are assumed to lie on lattice points; in addition, nano machine positions are assumed to be fixed. The Java-based simulator illustrated in \cite{Garralda2011}, called N3Sim, emulates a two-dimensional particle Brownian diffusion model, with 3D extensions in specific conditions. Propagation phenomena includes inertia forces and particle collisions. Given the particle emission process at different transmitters, it evaluates the evolution of the molecular concentration at each receiver located within an unbounded space. 
The receiving process consists of counting the particles located within a given region around the considered receiver. 
In \cite{Felicetti2012}, a Java package designed to simulate nano-scale communications in 3D, Biological and Nano-Scale communication simulator (BiNS), is presented.  The approach of BiNS is fine grained in that the position of each element is evaluated at each simulation step, and collisions are managed according to an elastic model. In the model, a number of receptors is distributed over the surface of each fixed or mobile nano machine. A nano machine receives a carrier when the latter hits one of the carrier-compliant receptors. In \cite{Felicetti2012}, BiNS is used for emulating a section of a lymph node and the information transfer within it, which happens between antibody molecules produced by the immune system during the humoral response.  In \cite{Felicetti201398}, the same authors present the version 2 of the BiNS package (BiNS2), which is able to simulate also partially inelastic collisions in bounded environment, such as the blood vessels. Finally, the BiNS2 package has been further enriched of an octree-based computation approach \cite{Felicetti2013172}, which uses a dynamic splitting of the simulated environment into cubes of different size in order to parallelize the simulation, so as to benefit of the multi-thread capabilities of modern multi-core computer architectures, and thus strongly reduce the simulation time. The BiNS2 simulation package has been used to carry out the simulations presented in section \ref{perf} of this work.
Finally, the work \cite{Akkaya2014163} presents a standardized simulation framework (High Level Architecture, HLA, defined in IEEE 1516), which is used to design and develop a distributed simulation tool for molecular communication, including the possibility to contemporary use different software tools to simulate different entities of the considered scenario.

\section{Protocol design}\label{design}

As anticipated in section \ref{intro}, the considered communication scenario consists of two fixed bio-nanomachines: one acting as receiver and control node (RX) and the other as transmitting node (TX). The communication happens by using two different types of molecules, which propagate by diffusion, modeled as Brownian motion \cite{Philibert2006}. The molecules transmitted by the TX node and representing the signal to be delivered are labeled as \textit{S}, and those transmitted by the control node RX to encode to control messages are labeled \textit{R}. The control signals are encoded in different patterns of bursts of the \textit{R} molecules. The proposed protocol consists of three main phases: the connection set up, the molecule delivery phase, and the connection tear down. The description of the protocol is organized as follows. First, in Section \ref{tx} we illustrate the framework used to transmit and receive control messages. Section \ref{enc} illustrates how control message are encoded.  In Section \ref{setup} we describe how connection set up is carried out, detailing the ranging procedure needed to identify the correct burst size of \textit{R} molecules, and the round trip time (RTT) estimation algorithm. The $S$ molecules delivery is illustrated in Section \ref{transfer}. The estimated RTT is used to decide when it is appropriate to send to the TX node a negative feedback in order to halve the current size of bursts of \textit{S} molecules, or to tear down the communication (Section \ref{closen}), by sending a stop signal. The control algorithms and the state machines for both the TX and RX nodes are presented. Table \ref{tabella_state} reports all parameters used in state machines.

\subsection{Transmission on the control channel}\label{tx}

In order to establish a reliable communication, the RX node needs to adopt a communication scheme which ensures uniqueness of transmitted messages and high probability of message reception. We assume that the RX node uses only one type of molecules \textit{R}, which are released in bursts, and each burst represents a binary symbol. The symbols are encoded through an on-off keying scheme, meaning that the symbol 1 is encoded through the transmission of a burst of \textit{R} molecules, whereas the symbol 0 is encoded by transmitting no molecules. The symbol duration is indicated by $T_S$. A symbol 1 is correctly received by the TX node if during the symbol time the number of received \textit{R} molecules is larger or equal than a threshold $\zeta_S$. Vice versa, a symbol 0 is correctly received by the TX node if during the symbol time the number of received \textit{R} molecules is lower than the threshold $\zeta_S$. Thus, defining as $n_{TX}(\tau)$ the number of molecules of type \textit{R} received at the time instant $\tau$ by the TX, this translates into the following condition for detecting a symbol 1 at time $t$ (the one for symbol 0 is dual):

\begin{equation}
	K(t)=\int_{t-T_S}^t n_{TX}(\tau) d\tau \geq \zeta_S.
\label{condition}
\end{equation}

The probability of receiving at least $\zeta_S$ \textit{R} molecules within a symbol time at TX, upon transmitting a burst of molecules of type \textit{R} by RX, has to be reasonably high to ensure communication reliability. Let us go a bit deeper on this. The probability density function (pdf) of receiving a molecule at time $t$ by a bio-nanomachine of radius $r_{TX}$ after the emission of a molecule at distance $d$ at time 0 is not known in a 3D space. In fact, only for a 1 dimension there is a closed form of this pdf (first hit time, see \cite{Nakano12d}). In addition, this closed-form pdf is evaluated by assuming that each hit of a molecule will imply an assimilation, which cannot alway be true, for instance when the number of receptors compliant with \textit{R} molecules covers just few percents of the overall bio-nanomachine surface. In any case, we can express such a probability density function $f_{TX}(t,d)$ as the conditional probability density function of receving a carrier at time $t$ conditioned by the event that a carrier has been received $f_{TX}(t,d | P_A(d))$ multiplied by the probability of assimilating a carrier $P_A(d)$ upon the transmission of a burst of \textit{R} molecules of size $Q$:

\begin{equation}
	f_{TX}(t,d)=f_{TX}(t,d | P_A(d)) \times P_A(d). 
\label{def}
\end{equation}

While $f_{TX}(t,d| P_A)$ can be obtained by simulation, we can elaborate a bit more on $P_A(d)$. In fact, the concentration at distance $d$, after $t$ seconds from the transmission in a 3D space of a burst of molecules of size $Q$ is known and equal to:

\begin{equation}
	c(t,d)= \frac{Q}{\left(4 \pi D t\right)^\frac{3}{2}} e^{\left(-\frac{d^2}{4D t}\right)},          
\label{conca}
\end{equation}

\noindent where $D$ is the diffusion coefficient, equal to $D=\frac{K_b T}{6 \pi \eta r_{c,rx}}$. $K_b$ is the Boltzmann constant, $T$ is the temperature in Kelvin degree, $\eta$ is the viscosity of the medium, and $r_{c,rx}$ is the radius of the molecules of type $R$. Starting from $c(t,d)$ and by using the first Fick's law \cite{Philibert2006}, it is possible to evaluate the flow of molecules as

\begin{equation}
	\textbf{J}(t,d)= -D \nabla c(t,d).          
\label{flux}
\end{equation}

By integrating $\textbf{J}(t,d)$ over the entire surface of the TX node (whose radius is $r_{TX}$) and over the time, it is possible to estimate the number of molecules that would cross the volume of the TX node if it would be virtual, that is as it would not cause a local perturbation to the system. However, the node TX is a sink for the molecules of type \textit{R}, and if we want to calculate the total number of absorbed \textit{R} molecules $A_{TX}(d)$, it is necessary to account for its presence by means of a correction function $\gamma(R_{TX})$, where $R_{TX}$ is the number of receptors present on the surface of the TX node and compliant with \textit{R} molecules. This function accounts for two different effects. The first one is that the TX node is a sink for type \textit{R} molecules, thus close to the surface of TX these molecules will experience a lower concentration. Since the first Fick's laws of diffusion (\ref{flux}) says that molecules tend to move towards the negative gradient of concentration, additional molecules will move towards TX and will be absorbed. The second effect to be modeled by $\gamma(R_{TX})$ is that the lower the number of \textit{R}-compliant receptors the lower the probability of an \textit{R} molecule close to the TX surface to be absorbed. Thus, $A_{TX}(d)$ will be given by:

\begin{equation}
A_{TX}(d) = \gamma(R_{TX})	 \int_{t=0}^{\infty} {\int_{S_{TX}} \textbf{J}(t,d) \bullet \textbf{n} dS_{TX}} dt,
\label{tt}         
\end{equation}

\noindent where $S_{TX}$ is the surface of the TX node, and $\textbf{n}$ is the unit vector orthogonal to each elementary surface portion $dS_{TX}$. 
When $d$ is at least one order of magnitude larger than $r_{TX}$, it is possible to assume that the spherical surface of the TX node can be replaced with a disc with the same area $S_{TX}=4 \pi r_{TX}^2$, whose center is located in the center of the TX node, and whose unit vector $\textbf{n}$ is aligned with the line connecting the centers of TX and RX nodes. In this case (\ref{tt}) may be approximated by 

 \begin{equation}
A_{TX}(d) \approx \gamma(R_{TX})	 \int_{t=0}^{\infty} {\int_{S_{TX}} \frac{Q d}{16 \left(\pi Dt \right)^\frac{3}{2} t} e^{\frac{-d^2}{4 D t}} dS_{TX}} dt =  \gamma(R_{TX}) Q \frac{4 \pi r_{TX}^2}{4 \pi d^2}  =  \gamma(R_{TX}) Q \left(\frac{r_{TX}}{d} \right)^2.   
\label{ntot}
\end{equation}

By looking to (\ref{ntot}), it is evident that:
\begin{itemize}
\item[-] the total number of assimilations scales linearly with the burst size $Q$. This is clearly reasonable for low numbers of absorbed molecules $A_{TX}(d)$, since for large values additional saturation phenomena enter into play (see also \cite{nakano13});
\item[-] the total number of molecules which flux through the TX surface, without accounting for the actual presence of the TX node, would be simply given by Q multiplied by the square ratio between the TX surface and the overall surface of radius $d$.
\end{itemize}

Thus, for low values of $A_{TX}(d)$, it is possible to easily evaluate the probability of assimilation as

\begin{equation}
P_A(d) =  \frac{A_{TX}(d)}{Q} = \gamma(R_{TX}) \left(\frac{r_{TX}}{d} \right)^2.   
\label{PA}
\end{equation}

Evidence of the correctness of the above mathematical derivation will be provided in section \ref{perf}, where theoretical curves will be compared with simulation results. Finally, as for the function $\gamma(R_{TX})$, we found experimentally that its behavior resembles that of the reaction rate in enzymatic kinetics \cite{nakano13,Keener08}, and, as expected, tends to saturate for very large values of the number of receptors $R_{TX}$. $C_1$ and $C_2$ are fitting coefficient, determined by simulation.

\begin{equation}
\gamma(R_{TX}) = \frac{C_1 R_{TX}} {C_2 + R_{TX}}  
\label{PA}
\end{equation}

Now, we can evaluate quantities relevant to reliability of communications. Let us define $P_{A,max}(d,T_S)$ as the maximum probability of having an assimilation at the TX node during a time interval equal to the symbol duration $T_S$, that is 

\begin{equation}
P_{A,max}(d,T_S) = \max \int_{t}^{t-T_S} f_{TX}(\tau,d) d\tau.   
\label{PAmax}
\end{equation}

This probability occurs when the TX node is perfectly synchronized with the RX one, that is TX is able to statistically capture the peak of assimilations during a symbol duration. Thus, by using the procedure illustrated in  \cite{Kuran201086}, in this condition the probability $P_C$ of correctly receiving a symbol (in this case a 1) when the RX node has sent a burst of size $Q$ can be calculated by

\begin{equation}
P_C(d,T_S) = \sum_{k=\zeta_S}^{Q} \binom{Q}{k} P_{A,max}(d,T_S)^k \left(1-P_{A,max}(d,T_S)\right)^{Q-k} = \cB\left(Q, P_{A,max}(d,T_S)\right),   
\label{prob_0}
\end{equation}

\noindent where $\cB(n,p)$ stands for the binomial distribution with parameter $n$ and success probability $p$.
Through an extensive numerical analysis, also taking into account false detection probabilities due to previous symbols transmissions (see again the detailed treatment in \cite{Kuran201086}), we have determined the values of the two parameters $\zeta_S$ and $T_S$ which are valid for an ample range of distances (about from $d$=20$\mu$m to $d$=70$\mu$m) and make the communication reliable. Clearly, when increasing the distance, it is necessary to increase the burst size $Q$ emitted by the RX node, as explained in the following section \ref{setup} and how it emerges from analysis of (\ref{prob_0}).

\subsection{Control messages encoding}\label{enc}

Once the values of the detection threshold and of the symbol duration have been defined, let us switch to the definition of control messages. We have defined three messages:
\begin{itemize}
\item[-]Message ``START'' to set up the connection,  through which the RX notifies the TX node to start emitting molecules; 
\item[-]Message ``STOP'' to tear down the connection,  through which the RX notifies the TX node to stop emitting molecules; 
\item[-]Message ``HALVE'', through which the RX notifies the TX node to halve the size of the current burst of molecules.
\end{itemize}

Thus, the control information consists of the delivery of messages, belonging to a source alphabet having a cardinality of 3.  We can assume that the probability of the START and STOP messages, respectively, are equal, since these messages delimit communication sessions. The HALVE message might either be sent multiple times during a communication session, or never. If we assume that the probability of the HALVE message is higher than those of START and STOP, as expected in the considered scenario, where the receiver may saturate frequently \cite{nakano13}, the resulting Huffman encoding results to be very simple, thus producing the following codewords: HALVE=``0'', STOP=``11'', and START=``10''. Given the particular communication environment based on particle diffusion, asynchronous transmission, and the use of the on-off keying, it was not possible to use codewords beginning with a symbol ``0''. Thus, for synchronizing the receiver, we made use of a line coding consisting of a further bit 1 attached to the Huffman codewords, obtaining HALVE=``10'', STOP=``111'', and START=``110''. We do not use ARQ techniques to acknowledge these messages, as illustrated in \cite{Leeson2014}, since it would need an additional molecule type to be used from TX to RX to carry these acknowledgments, different from $S$ molecules.

\begin{figure}[!ht]
\centering
\includegraphics[width=1\linewidth]{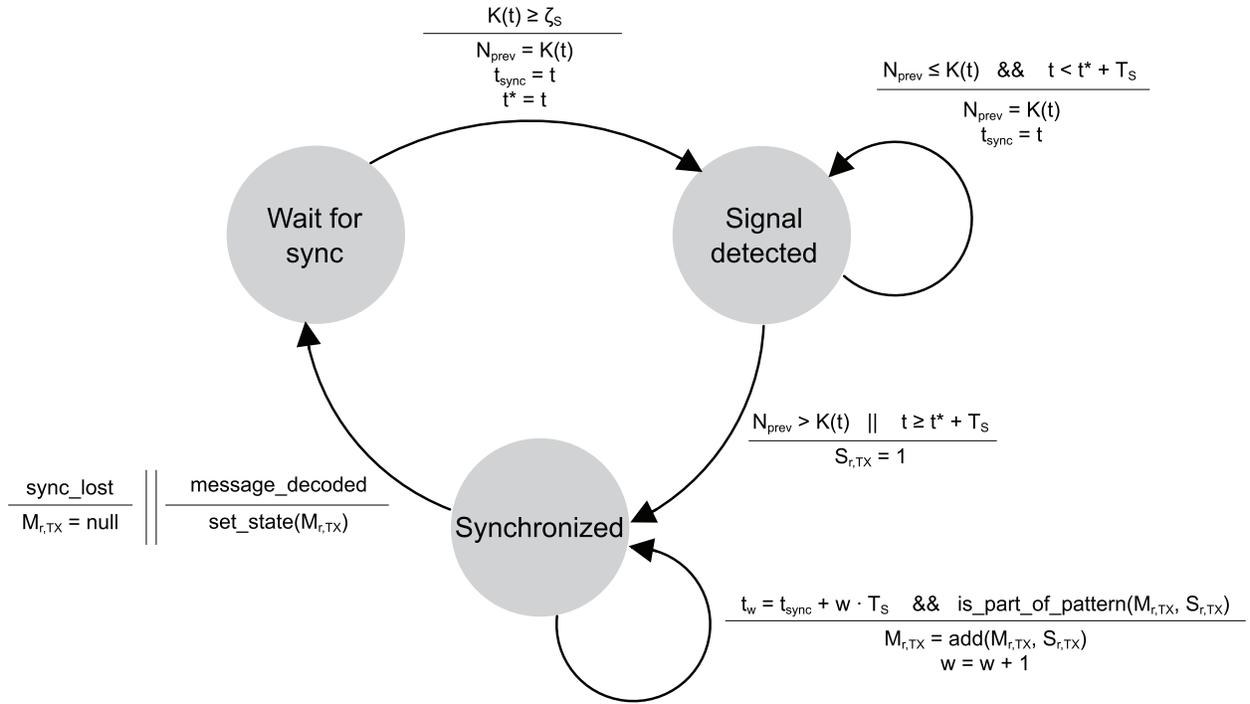}
 \caption{State machine for the synchronization algorithm.}
 \label{synchro}
\end{figure}

In order to establish a successful communication between different nodes placed at unknown distances, it is necessary to study a synchronization algorithm which, together with the detection procedure illustrated above, would be able to automatically synchronize \textit{without} any additional external reference signal.
In this regard, we have analyzed the twofold aspect of the signal detection and message decoding. The receiving node TX must determine whether the total amount of assimilated molecules during the symbol time, i.e. $K(t)$, is higher than a predefined threshold $\zeta_S$, in order to correctly decode a bit 1 or 0, as illustrated above.
The transmitted signals may have variable length due to the pulse spreading in the free space between RX and TX, so it is tricky for the receiving node to correctly decode consecutive messages. For these reasons, a synchronization phase is mandatory.
At rest, the TX node waits for the synchronization symbol (``wait for sync'' state). The procedure that we have designed is illustrated in Fig. \ref{synchro}. The value $t$ represents the current time. 
From the instant when the assimilated carriers in the last time window of length $T_S$ is higher than the threshold $\zeta_S$,  occurring at $t=t^{*}$, the node TX switches to ``signal detected'' state and periodically compares the current assimilation value with the one of the previous period; the value of such a period is equal to $T_S/20$, in order to filter out statistical oscillations due to randomize arrivals of new assimilations. If the number of assimilations increases, then the TX refresh both its synchronization time ($t_{sync}$) and the value of the last assimilation ($N_{prev}=K(t-T_S/20$)), in order to find the instant of the maximum assimilation (i.e. of $K(t)$) in the current time window $T_S$.
When the time window $T_S$ expires or the current assimilation value is lower than the previous one, the TX node switches to ``synchronized'' state and it assumes that the first symbol has been correctly decoded as 1. 

Clearly, the time shift between the detection of the signal (i.e. the first time in which condition (\ref{condition}) is verified, $t^{*}$) and the synchronization time $t_{sync}$ cannot be larger than $T_S$. In fact, this would imply that two consecutive 1s have been superposed, and consequently it would imply the detection of a single 1 instead of two consecutive 1s. This is more likely as the distance $d$ between the centers of TX and RX nodes increases, due to the behavior of the concentration of molecules over time $t$ and distance $r$ (\ref{conca}).

\begin{figure}[!ht]
\centering
\includegraphics[width=0.6\linewidth]{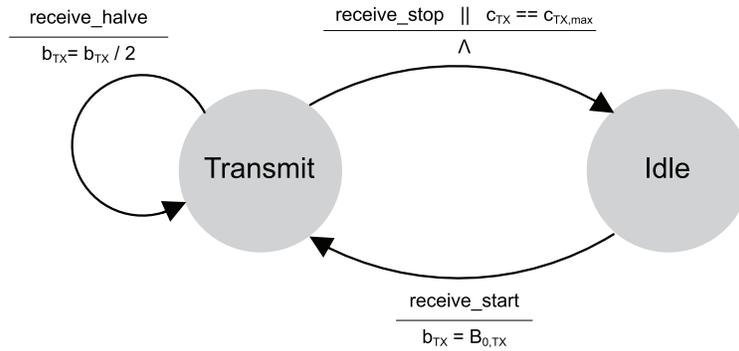}
 \caption{State machine for the TX node.}
 \label{FSM_TX}
\end{figure}

The synchronization time ($t_{sync}$) is used to enable a periodic reading of $K(t)$, with period $T_S$ starting form $t_{sync}$. In this way, each $T_S$ the TX node  checks if the current symbol is one of the expected symbols for the known messages, compliant with the current TX state (HALVE or STOP, see Fig. \ref{FSM_TX}) or not. If it is so, the current symbol will be attached to the tail of the previously assimilated sequence. Otherwise, it will be discarded, as shown in  Fig. \ref{synchro}.

Please note that the TX node will exit from the current state if the  synchronization is lost or even if the sequence matches with one of the known messages. In the former case, a wrong symbol on a given position of the sequence will lead to reboot the entire process, erasing the currently assimilated partial sequence. The latter case happens whenever the sequence of symbols matches with one of the known messages, so the internal state of the TX node is set as a function of the decoded signal.

\subsection{The connection set up}\label{setup}
As illustrated in the previous section, the probability to successfully set up a connection with the TX node depends on the value of the burst $b_{RX}$ of \textit{R} molecules used to send symbols by the RX node. 

\begin{figure}[!ht]
\centering
\includegraphics[width=1\linewidth]{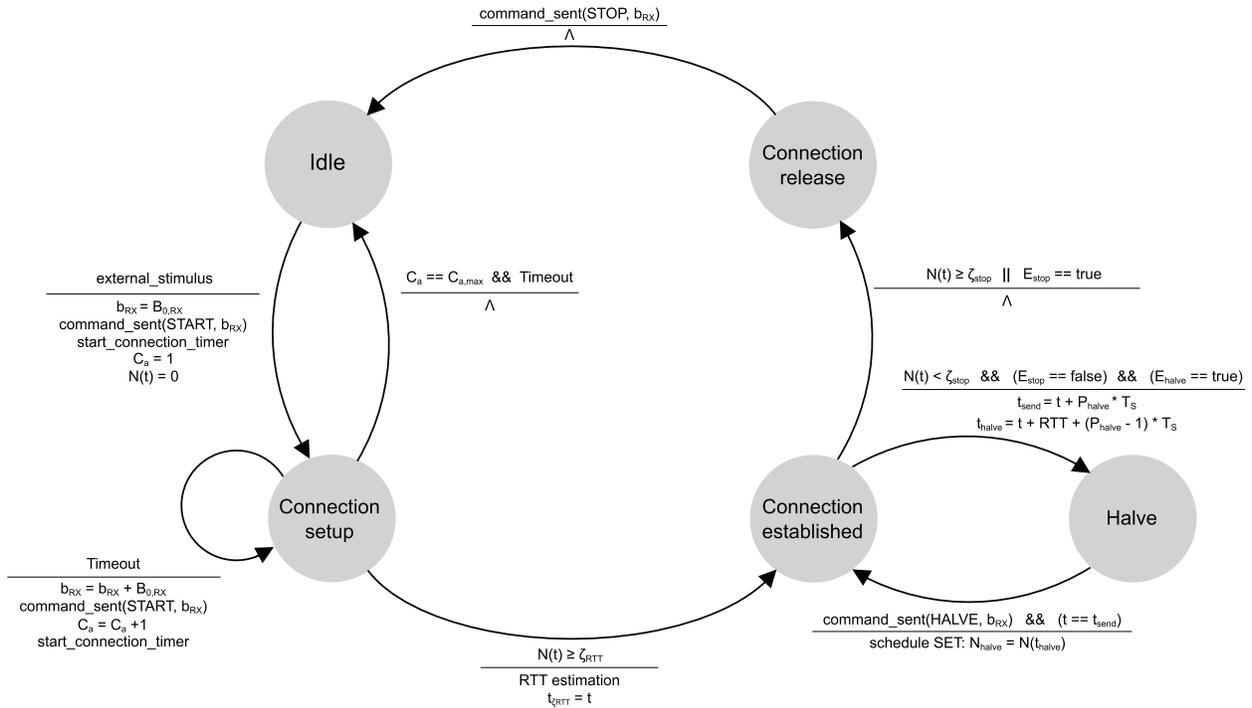}
 \caption{State machine for the RX node.}
 \label{FSM_RX}
\end{figure}

Initially, the control node RX is in the idle state. The complete state machine is illustrated in Fig. \ref{FSM_RX}. An external stimulus (e.g. a detection of a tumoral cells towards which trigger the drug delivery) may trigger a state change to ``connection setup'', where it emits the START signal as a train pulses, each one composed by $b_{RX}=B_{0,RX}$ carriers. The RX node assumes that the START signal is correctly received by the other node as soon as it senses its response. If that event does not happen within the timeout, the START signal will be sent again by using a larger burst for encoding each symbol equal to 1. At each attempt the size of the burst $b_{RX}$ is increased by a fixed quantity, that in our scheme is equal to $B_{0,RX}=1000$ molecules of type \textit{R}. The controller will try to send the START signal for a predefined number of times ($C_{a,max}$), then it will be back to the idle state, assuming that no TX nodes within the communication range associated with the maximum value of $b_{RX}=B_{0,RX} C_{a,max}$. This procedure closely resembles the ranging procedure used by a WiMAX terminal to initiate the communication with a WiMAX base station which has been detected in the communication range \cite{WiMAX}. Please note that having selected the values of $\zeta_S$ and $T_S$ which are valid for an ample range of distances $d$, this allows selecting the energy of the START signal (i.e. $(P_{start}-1) b_{RX}=2 b_{RX}$) which is more suitable in the current environment. 

\begin{figure}[!ht]
\centering
\includegraphics[width=0.9\linewidth]{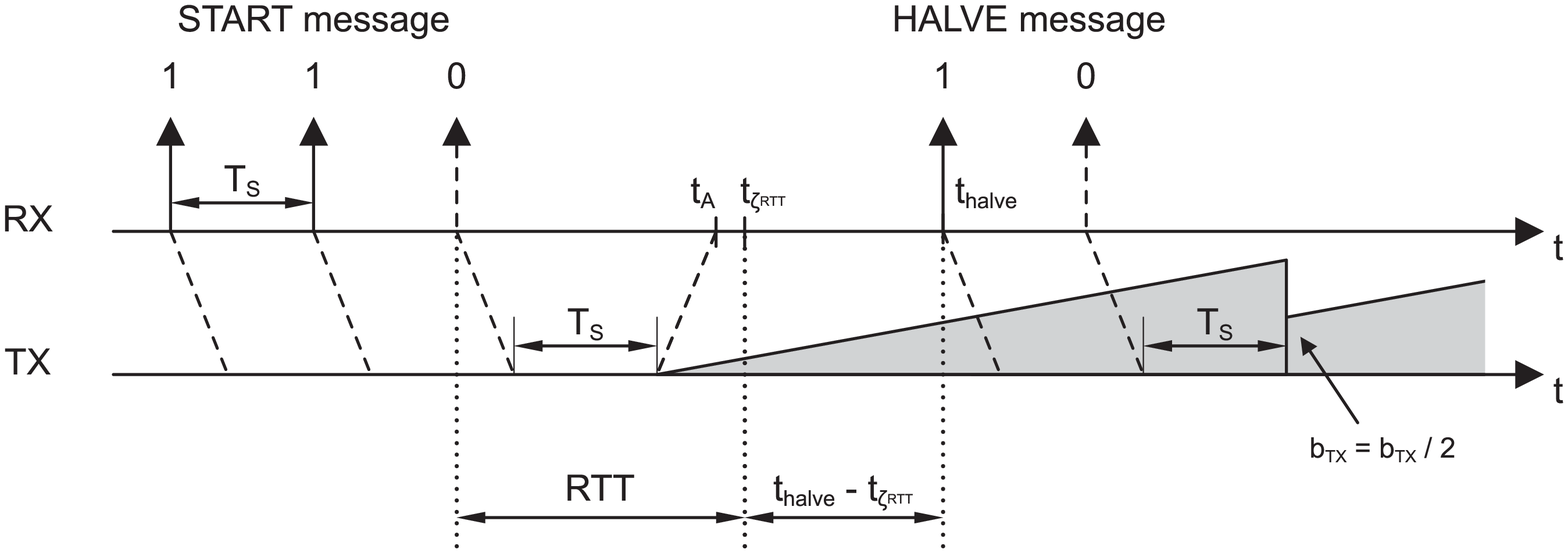}
 \caption{Temporal evolution of the protocol functioning.}
 \label{time}
\end{figure}

If the RX starts receiving molecules of type \textit{S}, it will wait until the reception of at least $\zeta_{RTT}$ type \textit{S} molecules to decide that the connection was established (switch to ``connection established'' state) . The time of this event is labeled as $t_{\zeta_{RTT}}$, and the RX node estimates the round trip time (RTT) equal to the time elapsed since the transmission of the last symbol of the START (i.e. the final 0) up to $t_{\zeta_{RTT}}$, as illustrated in Fig. \ref{time}, which shows the temporal evolution of the initial phase of the communication between RX and TX. $t_A$ represents the time at which the first $S$ molecule is absorbed by RX. This scheme to estimate the RTT is similar to the RTT-T one defined in \cite{Moore12a}.
 
\subsection{Molecules delivery}\label{transfer}

When the connection is set up, the two nodes have a different behavior. The TX node exhibits a very simple behavior, described by the finite state machine illustrated in Fig. \ref{FSM_TX}. Basically, upon the connection set up, the transmission burst is initialized to $B_{0,TX}=1$ molecule. These bursts of molecules are increased and released each $\Delta t$ by a fixed quantity, that in our scenario is fixed to 1 molecule. This choice is motivated by the fact that the TX node is not able to estimate the distance $d$ from the RX node and thus the optimal transmission rate \cite{nakano13}, since we decided to keep it simple and concentrate most of logic in the RX node. Thus, analogous to what does the TCP sender, which does not know in advance the maximum capacity of the network crossed by its packets, it ``probes'' the network by increasing the transmission rate. For the TCP, it happens first exponentially, and then, in the congestion avoidance, linearly with the RTT. In our protocol, for simplicity we designed the protocol to increase the release rate linearly, and assigned to the RX the task to throttle it when needed. On a large time scale, this means that after $ (n-1) \Delta t$ transmission intervals since the beginning, the total number of molecules $c_{TX}$ released by the TX node is equal to 

\begin{equation}
c_{TX} = \sum_{b_{TX}=B_{0,TX}}^{n B_{0,TX}} b_{TX}=\frac{n(n+1)}{2}B_{0,TX}=\frac{n(n+1)}{2},   
\label{prob_1}
\end{equation}

\noindent thus the number of molecules \textit{S} released in the surrounding space increases with a quadratic law. 

\subsection{Flow control and connection tear down}\label{closen}

Since the number of absorbed molecules is proportional to the number of released ones (see Section \ref{tx}), this means that also the number of  molecules received by RX should follow a quadratic law. However, when the number of molecules is very large, the RX could undergo to a saturation condition, which implies that its receptors are no more able to chemically react with all these molecules, and the curve $N(t)$ describing the number of received molecules start deviating from a quadratic law. This is the case in which the efficiency of the communication, defined as the ratio between the absorbed molecules by RX and those released by TX, start decreasing, and it is necessary to send a control signal.

The control node senses the channel periodically with period $T_w$ in order to decide if it is necessary to send a control message to limit the emission at the TX node in two alternative ways:  
\begin{itemize}
\item[-] by sending the HALVE signal, if (i) the number of received molecules is large enough (i.e. larger than $\zeta_{halve}$) to be able to reliably estimate the scaling coefficient $a$ for the quadratic law governing the reception process, and (ii) the number of currently received molecules significantly deviate, beyond a tolerance margin $\beta$, from the expected one on the basis of the previously calculated value of $a$;
\item[-] by sending the STOP signal if either the total number of received molecules is currently larger than the target $\zeta_{stop}$, or if its estimated value, at the end of the transmission of the STOP signal plus the estimated RTT value, could be reasonably larger than the target value, taking into account the number of molecules in-flight at the time of decision.
\end{itemize} 

\begin{figure}[!ht]
\centering
\includegraphics[width=0.7\linewidth]{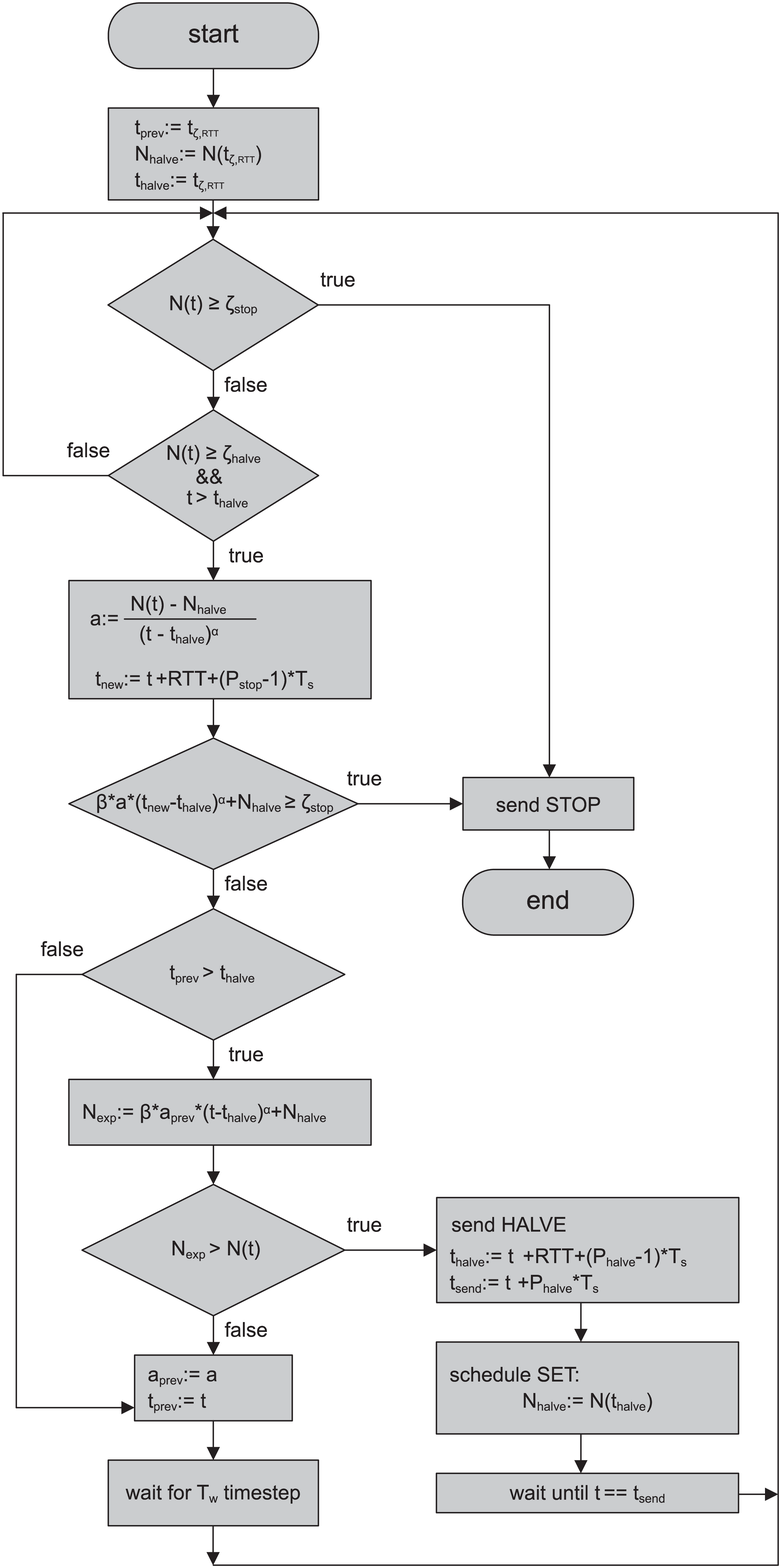}
 \caption{RX control algorithm executed in the ``Connection established'' state.}
 \label{parabolic}
\end{figure}

The detailed algorithm is reported in Fig. \ref{parabolic}. When one of these conditions is verified, a control signal will be sent.
For the first case, if it detects that the total number of assimilations does not exceed the stopping threshold and only the HALVE estimation ($E_{halve}$) is verified, then it decides to send the HALVE signal $(command\_sent(HALVE,b_{RX}))$. As soon as the time needed to send that message is elapsed, the RX node goes back to ``connection established'' state in order to sense again the channel.

If the number of assimilations exceeds the threshold $\zeta_{stop}$, or the stop estimation condition ($E_{stop}$) is verified, the RX switches to the ``connection release'' state. The stop estimation condition estimates the number of carriers that will be likely assimilated during the time needed for transmission, propagation and decoding of that signal.  
This estimation allows sending a STOP even if the halve condition is verified as well. In fact, sending of consecutive HALVE and STOP commands is avoided, since it would cause an excessive molecules assimilation within the time needed to send both commands. This procedure avoids receiving an excessive number of molecules. Since the RTT is estimated when the first $\zeta_{RTT}$ molecules are received, the stop estimation condition does not take into account the tail of the emitted molecules. Hence, these molecules provide a suitable safety margin which avoids tearing down the connection without having achieved the target. In the worst case, if the STOP signal is sent too early, the RX node can set up another connection to complete the delivery of the desired amount of \textit{S} molecules.

As soon as the STOP signal has been transmitted, the node switches to the ``idle state''.

When the TX node receives an HALVE message, it simply halves the burst size $b_{TX}$, and then continues to increase it by $B_{0,TX}$ each $\Delta t$ as before, as in the ``Fast recovery'' procedure of the TCP Reno. The TX node re-enters in the ``idle state'' when it receives and correctly decodes a STOP signal, or has delivered a maximum, pre-defined number of molecules ($c_{TX,max}$). This additional check ensures  that, if one or both the  bio-nanomachines would be slightly mobile, they could also lose the connectivity during the transfer phase, and this would preserve part of stoked molecules from being dispersed in the environment (soft state management).

When the HALVE message is sent, we have included in the design of the control algorithm of RX a time interval to perform again the estimation of the reception process before estimating if another HALVE, a STOP, or none of them would be necessary.

If both conditions are not verified, the quadratic coefficient $a$ is determined, as briefly anticipated above. This coefficient is used to estimate the assimilations at the next control time, that is $t+T_w$. Intuitively, $a$ is obtained from the ratio between the number of assimilations in the last observed time window ($N(t)-N_{halve}$) and the  time window itself ($t-t_{halve}$). The coefficient $\alpha=2$ represents the order of the polynomial and $N_{halve}$ is the amount of assimilations after the round trip propagation of the HALVE message (note that if no HALVE  message has been sent, $N_{halve}=N(t_{\zeta_{RTT}})$).
From the RTT and the time needed to send the stop message ($(P_{stop}-1)T_S$), it is possible to obtain the minimum time ($t_{new}$) at which to observe the effects that the STOP signal would produce on the assimilations of the RX node.

The sum of both estimated ($\beta a (t_{new}-t_{halve})^{\alpha}$) and past assimilations ($N_{halve}$) is then compared with the stopping threshold $\zeta_{stop}$. If such a value is higher or equal to that threshold, the STOP signal is sent, otherwise it means that at time of the next $t_{new}$ the expected assimilations likely will not reach the target $\zeta_{stop}$ and the emission should not be stopped.
If this is the case, the algorithm goes on to check if it is necessary to reduce the emission process due to saturation condition of its surface receptors. 

\section{Performance evaluation}\label{perf}

The performance evaluation of the system has been carried out by using the BiNS2 simulator. The main simulation parameters, together with their description and values, are reported in Table \ref{tabella}.

\begin{table}[t]
\caption{Parameters of the protocol state machines.}
\centering
\scalebox{0.80}{
\begin{tabular}[t]{p{3.2cm}|p{7.7cm}|p{2.6cm}|p{1.1cm}}
Symbol & Description & Type	& Entity\\ \hline
$T_{S}$	& Symbol time &	state variable &	TX, RX\\
$\zeta_{S}$	& Assimilation Threshold & state variable &	TX\\
$t_{sync}$	&Synchronization time &state variable &	TX\\
$t*$	& Provisional synchronization time 	& state variable	& TX\\
$N(t)$	&Total current assimilations	of $S$ molecules& state variable	& RX\\
$K(t)$	&Assimilations of $R$ molecules occurred in the last $T_{S}$	& state variable	& TX\\
$N_{prev}$	& Provisional number of molecules assimilated in the last $T_{S}$& state variable	& TX\\
$M_{r,TX}$	& Received message	& state variable	& TX\\
$S_{r,TX}$ &	Received symbol	& state variable	& TX\\
$t_{w}$	& Time instants in which a symbol is decoded	& state variable	& TX\\
$w$	& Counter of received symbols 	& state variable	& TX\\
$b_{RX}$	& Current burst size (node RX)	& state variable	& RX\\
$b_{TX}$ & Current burst size (node TX)	& state variable	& TX\\
$B_{0,RX}$	& Initial burst size (node RX)	& state variable	& RX\\
$B_{0,TX}$ &	Initial burst size (node TX)	& state variable	& TX\\
$c_{TX}$ &	Total number of $S$ molecules sent by TX& state variable	& TX\\
$c_{TX,max}$ &	Maximum number of $S$ molecules that TX	can send in a single session & protocol parameter& TX\\
$C_{a}$	& Connection attempts	& state variable	& RX\\
$C_{a,max}$ &	Max connection attempts	& state variable	& RX\\
$\zeta_{RTT}$	& Assimilation threshold for RTT estimation	& protocol parameter	& RX\\
$t_{\zeta_{RTT}}$	& Connection established time	& state variable	& RX\\
$t_{A}$	& Time instant of the first carrier assimilation	& state variable	& RX\\
$\zeta_{stop}$ &	Target number of S molecules to assimilate	& protocol parameter	& RX\\
$E_{stop}$ &	Estimation to send STOP signal	& state variable	& RX\\
$E_{halve}$ &	Estimation to send HALVE signal	& state variable	& RX\\
$t_{send}$	& Time instant when the transmission of a pattern ends& state variable	& RX\\
$P_{start}$ &	Amount of symbols for HALVE signal	& protocol parameter	& RX\\
$P_{halve}$ &	Amount of symbols for HALVE signal	& protocol parameter	& RX\\
$P_{stop}$ &	Amount of symbols for STOP signal	& protocol parameter	& RX\\
$t_{halve}$	& Time instant when the effect of halve signal should reach RX	& state variable	& RX\\
START	& START message	& protocol message& RX\\
HALVE	& HALVE message	& protocol message& RX\\
STOP	& STOP message	& protocol message& RX\\
$N_{halve}$ &	Amount of assimilations after the round trip propagation of the HALVE message	& state variable	& RX\\
external\_stimulus	& External stimulus that activates the node	& event	& RX\\
sync\_lost	& Synchronization lost	& event	& TX\\
message\_decoded	&The received pattern has been correctly decoded into $M_{r,TX}$& event	& TX\\
set\_state($M_{r,TX}$)	& Sets the state according to the received message $M_{RX}$	& action	& TX\\
is\_part\_of\_pattern(a,b)	& Checks if the current symbol b is one of the expected ones on the partial pattern a & action	& TX\\
command\_sent(cmd,$b_{RX}$)	& Send the command cmd=\{START|STOP|HALVE\}, each symbol is composed by $b_{RX}$ carriers	& action	& RX\\
start\_connection\_timer	& Starts the timer to monitor the connection setup	& action	& RX\\
Timeout	& The time to establish a connection has expired	& event	& RX\\
RTT estimation	& Estimates the Round Trip Time	& action	& RX\\
receive\_halve	& The node has received the HALVE signal	& event	& TX\\
receive\_stop	& The node has received the STOP signal	& event	& TX\\
receive\_start	& The node has received the START signal	& event	& TX\\
\end{tabular}
\label{tabella_state}
}
\end{table}

\begin{table}[t]
\caption{Simulation parameters}
\centering
\begin{tabular}{l|l|l}
Symbol & Description & Value\\ \hline
$dt$ & Simulation time step & 20 $\mu s$ \\
$T$ & Temperature & 310 K \\
$e$ & Coefficient of restitution & 0.9 \\
$\eta$ & Viscosity & 0.0011 $Kg \times (m s)^{-1}$\\
$\alpha$ & Grow factor for assimilations in RX & 2 \\
$\beta$ & Tolerance factor & 0.95 \\
$T_S$ & Symbol time & 10 s \\
$r_{RX}$ & Radius node RX & 2.5 $\mu m$ \\
$r_{TX}$ & Radius node TX & 2.5 $\mu m$ \\
$R_{RX}$ & Amount of surface receptors (node RX) & 10000 \\
$R_{TX}$ & Amount of surface receptors (node TX) & 10000 \\
$r_{c,rx}$ & Radius emitted molecules (type \textit{R}) & 3.5 nm \\
$r_{c,tx}$ & Radius emitted molecules (type \textit{S}) & 1.75 nm \\
$r_{r,rx}$ & Receptor radius (node RX) & 8 nm \\
$r_{r,tx}$ & Receptor radius (node TX) & 4 nm \\
$T_{traff}$ & Trafficking time \cite{Lauffenburger1993} & 200 $\mu s$ \\
$\zeta_S$ & Assimilation Threshold (node TX) & 34 molecules\\
$\Delta t$ & Emission time (node TX) & 20 ms \\
$Timeout_{rx}$ & Timeout before retransmission (node RX) & 54 s \\
$\zeta_{halve}$ & Assimilation Threshold for HALVE signal (node RX) & 250 molecules \\
$\zeta_{stop}$ & Assimilation Threshold for STOP signal (node RX) & 10000, 20000 molecules \\
$START$ & Signal pattern: START & 110 \\
$HALVE$ & Signal pattern: HALVE & 10 \\
$STOP$ & Signal pattern: STOP & 111 \\
$d$ & Simulated distance ($\mu m$) & 26.5, 35.4, 44.2, 53.0, 61.9 \\
$B_{0,RX}$ & Initial burst (node RX) & 1000 \textit{R} molecules \\
$B_{0,TX}$ & Initial burst (node TX) & 1 \textit{S} molecule \\
$\zeta_{RTT}$ & Assimilation Threshold for RTT estimation (node RX) & 5 molecules\\
$T_w$ & Waiting time for parabolic estimation (node RX) & 0.2 s \\
\end{tabular}
\label{tabella}
\end{table}

\begin{figure}[!ht]
\centering
\includegraphics[width=1\linewidth]{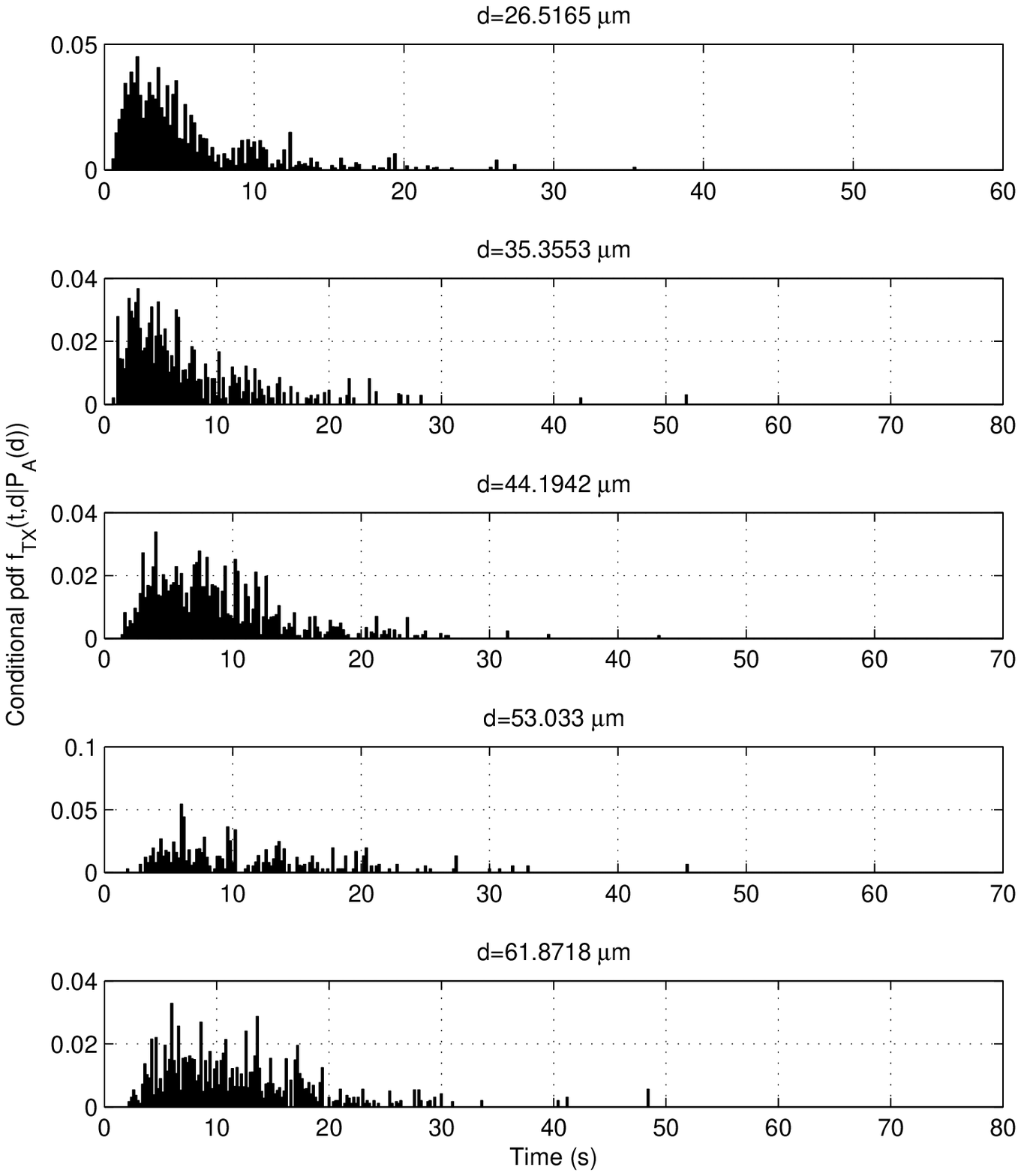}
 \caption{The conditional probability density function $f_{TX}(t,d|P_A(d))$ versus time, as a function of the distance $d$ between RX and TX centers.}
 \label{list}
\end{figure}

First, let us analyze the physical characteristics of the channel, which lead to the choice of the values of $T_S$=10 s and $\zeta_S$=34 molecules used in the simulator. Fig. \ref{list} illustrates the probability density function of assimilation time conditioned to the event of having an assimilation at the TX node, defined by (\ref{def}) in section \ref{tx}. As anticipated, it is evident that the larger the distance from the emission point, the larger the spreading of the ``signal''. This translates into the fact that that the symbol time practically covers all the useful time spread of the signal for $d$ equal to about 20 $\mu m$ ($P_{A,max}(26 \mu m,T_S)=0.889$), whereas it decreases for larger values, down to $P_{A,max}(62 \mu m,T_S)=0.6613$ for $d$ equal to about 60 $\mu m$. 

The trend of change of the probability density function with the distance $d$ is determined by both equation (\ref{conca}) which models the concentration of molecules at distance $d$ and the reception process, which is affected by the number and distribution of receptors on the TX surface.

\begin{figure}[!ht]
\centering
\includegraphics[width=0.7\linewidth]{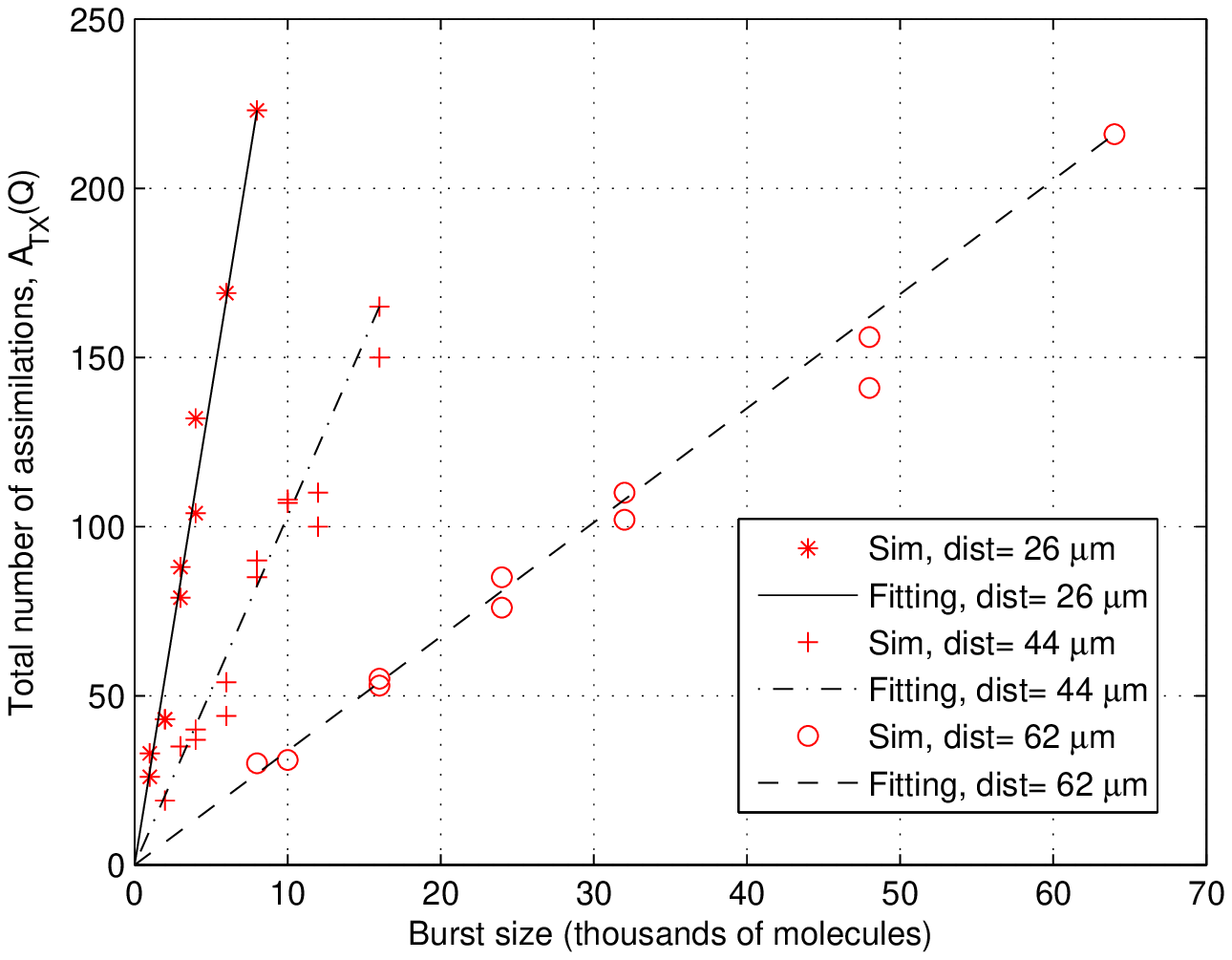}
 \caption{The total number of assimilations $A_{TX}(Q)$ versus bust size, for $d$= 26, 44, and 62 $\mu m$; $R_{TX}=10000$.}
 \label{f1}
\end{figure}

\begin{figure}[!ht]
\centering
\includegraphics[width=0.7\linewidth]{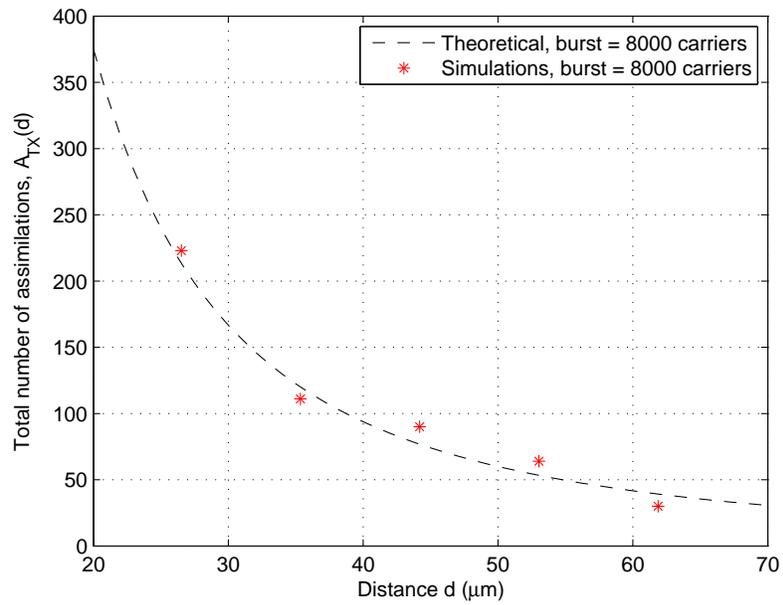}
 \caption{The total number of assimilations $A_{TX}(d)$ versus the distance $d$ between RX and TX centers: simulations and theoretical model for a burst of 8000 molecules, $R_{TX}=10000$.}
 \label{f2}
\end{figure}

\begin{figure}[!h7]
\centering
\includegraphics[width=0.7\linewidth]{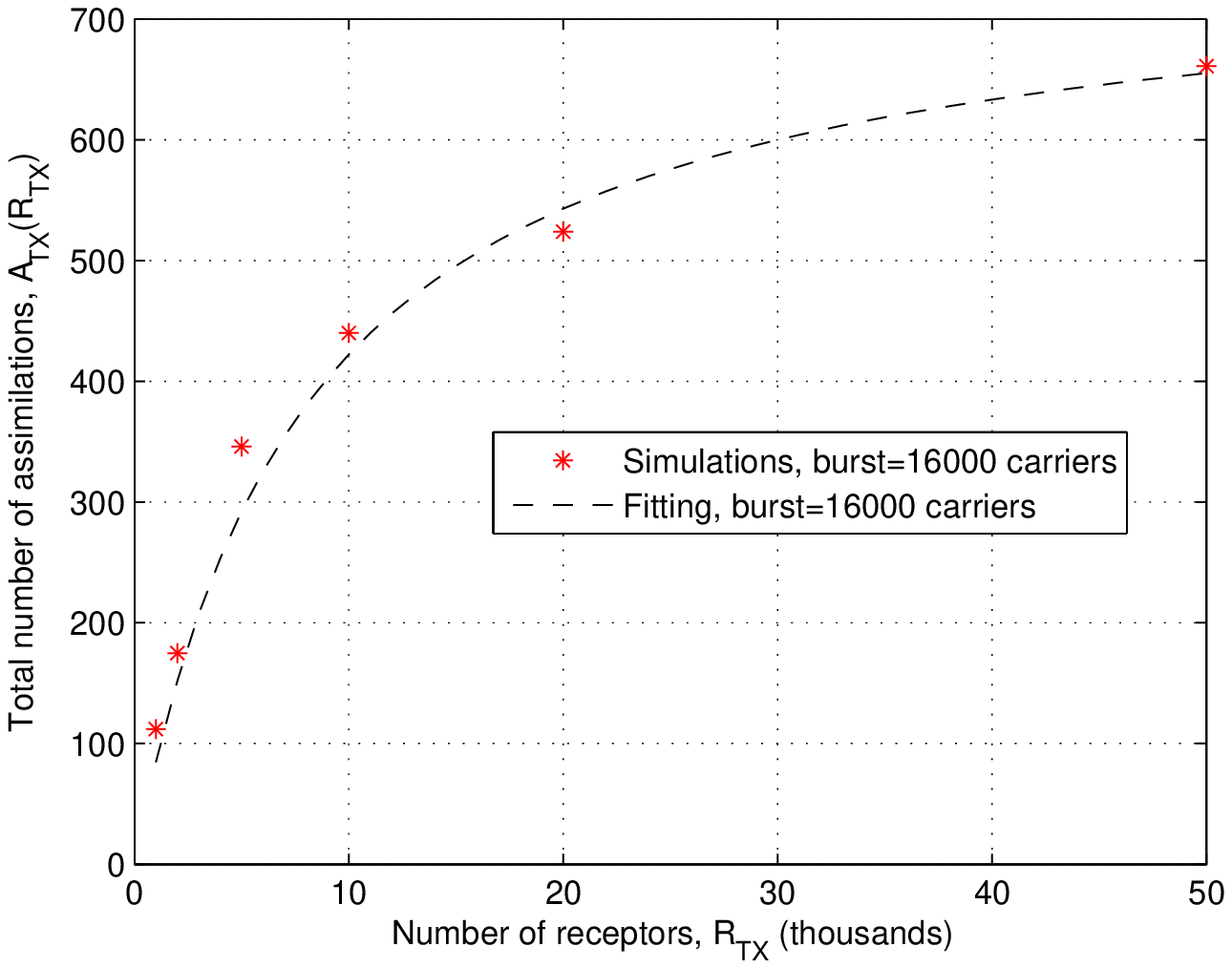}
 \caption{The total number of assimilations $A_{TX}(R_{TX})$ versus the number of receptors $R_{TX}$ of TX for $d=26 \mu m$: simulations and theoretical model for a burst of 16000 molecules.}
 \label{f3}
\end{figure}

Fig. \ref{f1} shows the dependency of the number of assimilations, $A_{TX}(Q)$, versus burst size. As already anticipated in section \ref{tx}, when the saturation condition does not hold, there is a perfect linear scaling between the number of emitted molecules at RX and those received at TX. The fitting curve can be easily evaluated a priori by using the estimation  provided for $A_{TX}(d)$ in (\ref{ntot}). The excellent agreement between the simulations and the theoretical curve can be observed in Fig. \ref{f2}. Finally, Fig. \ref{f3} shows the dependency of  the number of assimilations, $A_{TX}(R_{TX})$, versus the number of node receptors, $R_{TX}$. The values of the coefficient used in the fitting curve have been obtained numerically, and are equal to $C_1=5.344$ and $C_2=8000$ receptors. Thus the fitting equation used in the above figures and derived from (\ref{ntot}) is 

\begin{equation}
A_{TX}(R_{TX},d,Q) = \frac{C_1 R_{TX}}{C_2+R_{TX}}Q\left(\frac{r_{TX}}{d}\right)^2=5.344\frac{ R_{TX}}{8000+R_{TX}}Q\left(\frac{r_{TX}}{d}\right)^2.   
\label{fitt_A}
\end{equation}

Now, it is essential to consider a fundamental aspect necessary for implementing the protocol. During simulations, we realized that the $S$ molecules, sent by the TX, during their diffusion through the propagation medium, can obstruct the propagation of the feedback $R$ molecules. In more detail, if the size of the feedback molecules is smaller or equal to the size of the forward ones, which are much more numerous, $R$ molecules are bounced back and cannot reach the TX node. In other words, the forward molecules \textit{S} form a sort of wall for the feedback molecules, which, obeying to the first Fick's law, move towards the negative gradient of the concentration, and thus far from TX. In order to avoid this disrupting effect, we had to use a larger size and mass for the feedback molecules, as reported in Table \ref{tabella}.  In this way, they are much less affected by (partially inelastic) collisions with smaller $S$ molecules, and can reach the TX node.

Now, let us consider the performance of the protocol itself, that is throughput, efficiency, and amount of overhead due to the control messages. For these quantities, we have just a few initial results, which however capture very well the system dynamics. Note that the initial burst size at node TX is equal to 1 $S$ molecules, and then its size is regulated by the flow control driven by the receiver RX. The same regulation does not modify the burst of $R$ molecules after a suitable size is determined. 

\begin{figure}[!ht]
\centering
\includegraphics[width=1\linewidth]{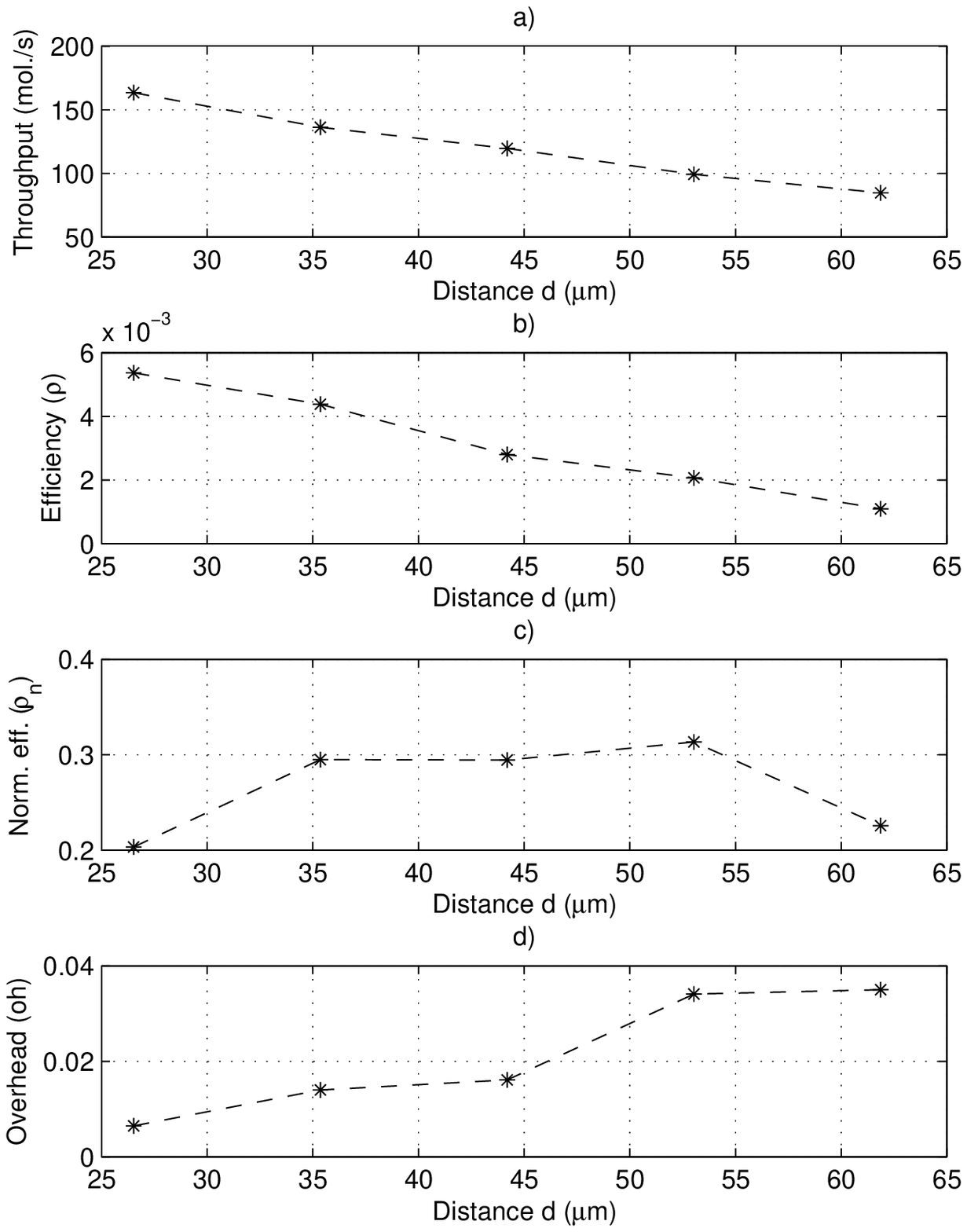}
 \caption{a) Throughput $thr$ versus distance $d$, b) efficiency $\rho$ versus distance $d$, c) normalized efficiency $\rho_n$ versus distance $d$, and d) overhead  $oh$ versus distance $d$.}
 \label{prot}
\end{figure}

We define the throughput (\textit{thr}) as the number of target molecules, $\zeta_{stop}$, divided by the time needed to deliver them ($T_D$), i.e. $thr=\zeta_{stop}/T_D$, without including the time contribution due to the failed ranging attempts. For instance, for $d=26\mu m$ the optimal burst size is 2000, for $d=35\mu m$ the optimal burst size is 3000, for $d=44\mu m$ the optimal burst size is 6000, for $d=53\mu m$ the optimal burst size is 10000, and for $d=62\mu m$ the optimal burst size is 16000.  Fig. \ref{prot}.a shows the throughput $thr$ as a function of the distance $d$. The behavior of the throughput is decreasing with $d$, which is an expected results. In fact, since for a given number of transmitted molecules of type \textit{S}, the fraction of them received by an RX node at distance $d$ decreases quadratically with $d$ (see also (\ref{ntot})), consequently the time needed to retrieve a fixed amount of them ($\zeta_{stop}$) increases as well, and thus the throughput decreases.

We define the efficiency ($\rho$) as the ratio between the number of target molecules at the RX node,  $\zeta_{stop}$, and the total number of molecules emitted by the TX node ($c_{TX}$), that is $\rho = \zeta_{stop} / c_{TX}$. Due to the considerations done while commenting the throughput performance figure, also in this case the performance worsens with $d$, as shown in Fig. \ref{prot}.b. In fact, when increasing the distance, the amount of molecules that the RX is able to intercept decrease with the square of the distance $d$, as shown in Fig. \ref{prot}.b. A further consideration is that the values of $\rho$ are in the range of $10^{-3}$, which is definitely too low. However, the pure efficiency is not a fair performance measure, since a receiver located at distance $d$ cannot capture \textit{all} transmitted molecules due to the intrinsic characteristics of the diffusion process, which governs the signal propagation. In order to provide a more complete set of parameters illustrating the actual system performance, we have introduced a normalized efficiency value, which refers uniquely to quantities collected at the receiver site. This performance figure is defined as $\rho_n= \zeta_{stop} /  A_{TX}(R_{TX},d,c_{TX})$. The value $A_{TX}(R_{TX},d,c_{TX})$ allows considering only the molecules that could be actually absorbed by the receiver, by using (\ref{fitt_A}). The obtained values give a more realistic view of the receiver performance, since $\rho_n$ is not affected by the molecules that propagate far away from the receiver and, therefore, cannot be received. Fig. \ref{prot}.c shows the values of $\rho_n$ as a function of the distance $d$. These efficiency values are much more sounding, since range between 20\% and 30\%, which are values typical also of widespread wireless network protocols, such as slotted Aloha \cite{Tanenbaum02}. The $\rho_n$ increases with the distance for $d$ passing from 26$\mu m$ ($\rho_n\approx20\%$) to 35$\mu m$ ($\rho_n\approx30\%$), and remains nearly flat up to $d=53 \mu m$.  This behavior, can be explained by considering that for $d\approx 26 \mu m$ the TX and RX are so close, with respect to the RTT and symbol time $T_S$, that when the RX estimates that it is necessary to send the STOP signal, it is a late estimation (see Fig. \ref{parabolic}). Instead, for larger distances, the larger propagation time slow down the process, and the control becomes more effective, and the protocol tends to saturate to the maximum values of the efficiency, which results from these tests equal to about 30\%. For $d\approx62 \mu m$, the normalized efficiency decreases to $\rho_n\approx22\%$. This is due to an opposite effect: the rate of increase of the transmitted molecules by TX is so large (see equation (\ref{prob_1})), that small errors in the estimation of the $E_{stop}$ condition are mapped into a decreased normalized efficiency.

Finally, we define the overhead (\textit{oh}) as the ratio between the total number of emitted \textit{R} molecules and the total number of emitted \textit{S} molecules. In more detail, $oh$ is given by

\begin{equation}
oh(d) = \frac{B_{0,RX} \left(\sum_{i=1}^{C_a}{i(P_{start}-1)} + P_{stop} C_a + n_{halve} C_a (P_{have}-1)    \right)}{c_{TX}}, 
\label{oh}
\end{equation}

\noindent where $n_{halve}$ is the number of HALVE messages sent during the simulation, and $P_{start}$, $P_{have}$, and $P_{stop}$ account for the number of bits trasmitted (see Table \ref{tabella_state} and Table \ref{tabella}). In this case, even if the total number of molecules transmitted by the TX nodes increases with the distance $d$ to reach the same target number at node RX, as explained just above, the particular ranging procedure developed for this protocol causes increasing the number of type \textit{R} molecules used to control the connection so much that the overhead slightly increases with the distance when passing from $d=44 \mu m$ to $d=53 \mu m$, and then remains nearly constant. However, the maximum value is just few percents, as shown in Fig. \ref{prot}.d, thus the protocol overhead is acceptable.

In addition, we have compared the system performance for two different values of the number of target molecules to be received by RX, $\zeta_{stop}$ equal to 10000 and 20000, for $d=26 \mu m$, reported in Table \ref{table2}. The main comment is that all performance figures improve when  $\zeta_{stop}$ increases. This can be explained as follows. As for the efficiency, since the distance is quite small, for $\zeta_{stop}=10000$ the RX would trigger the STOP signal very early, but it has to wait that the total number of received molecules reaches the threshold $\zeta_{halve}$ before sending any control message. Instead, when $\zeta_{stop}$ increases, this condition does not hold anymore, and the STOP signal is sent well after the time instant in which $\zeta_{halve}$ is reached. Thus, even if the numerator of the efficiency performance figures ($\rho$ and $\rho_n$) doubles, the denominator does not, and it only slightly increases, so that the efficiency increases for larger $\zeta_{stop}$. As for the throughput, it benefits from the quadratic profile of both emissions and assimilations, and thus throughput is constantly increasing during all the transfer phase. Finally, obviously also the overhead improves, since the larger number of molecules at the denominator is not compensated by a larger number of control molecules at the numerator.

Finally, additional preliminary results indicate that if the number of receptor decreases and the trafficking time $T_{traff}$ \cite{Lauffenburger1993} increases, the net effect is that the saturation is reached earlier, and a number of HALVE messages are sent before the completion of type \textit{S} molecules delivery.

\begin{table}[t]
\caption{Comparison for different values of $\zeta_{stop}$, $d=26\mu m$}
\centering
\begin{tabular}{l|l|l|l|l}
$\zeta_{stop}$ & Throughput & Efficiency ($\rho$) & Normalized efficiency ($\rho_n$)& Overhead\\ \hline
10000 & 163.4 molecules/s & 0.0054 & 0.2046 & 0.0064\\
20000 & 274.7 molecules/s  & 0.0079 &  0.2993 & 0.0047\\
\end{tabular}
\label{table2}
\end{table}

\section{Conclusion}\label{conc}

In this paper, we designed a communication protocol using molecular communications among bio-nanomachines. To the best of our knowledge, this is the first attempt to design a complete communication protocol using molecular communications.

 We first designed the control scheme at the physical layer, identifying a number of trade-off about system parameters (namely the symbol duration and the detection threshold) which allows using the system  for different distances of the bio-nanomachines, which are able to auto-configure and do not need any external intervention. Then, we proposed a finite state machine for both the transmitter and the receiver node, by borrowing functions and ideas from the well-known TCP Reno, and namely the congestion avoidance probing feature and the fast recovery. Due to the large delay which characterizes molecular communications, using explicit acknowledgments is not suitable, so we leverage on the results of previous research and adopt negative acknowledgments. In addition, we propose a number of control actions, which have been designed to prevent improper or inefficient protocol operation.

An additional contribution, emerged during the initial simulation set up, is the finding that in order to design an effective communication protocols, the molecules which are less numerous have to be a bit larger and with a larger mass.

Our ongoing and future work includes a complete performance evaluation of the protocol through simulations under a variety of different values for main simulation parameters, and in particular trafficking time and number of receptors. In addition, an additional goal is to extend the protocol in a multi-access environment, in which there are multiple TX and RX nodes, which interfere each other.

Finally, it is our aim to investigate specific application scenarios (e.g., drug delivery) considering application dependent constraints (e.g., specific spatial distributions of senders and receivers, flow/drift in the environment, a limited number of molecules stored in each bio-nanomachine).

\section{Acknowledgment}
We thank the anonymous reviewers for their helpful comments and suggestions that helped us largely improve the paper. This work was supported in part by Fondazione Cassa di Risparmio di Perugia, project code 2013.0055.021 RICERCA SCIENTIFICA E TECNOLOGICA.

\bibliographystyle{IEEEtran}
\bibliography{IEEEabrv,Femminella2}

\end{document}